\documentclass[journal]{IEEEtran}

\ifCLASSINFOpdf

\else

\fi

\usepackage{multirow}
\usepackage{pdflscape}
\usepackage{amssymb}
\usepackage{amsmath}
\usepackage{amsfonts}
\usepackage{graphicx}
\usepackage{xcolor}
\usepackage{cite}
\usepackage{subcaption}
\usepackage{mathtools}
\usepackage[utf8]{inputenc}
\usepackage{mathtools,amsmath}

\begin{document}

\title{Enhancing the Performance of Low Priority SUs Using Reserved Channels in CRN}

\author{Ahmed T. El-Toukhy~\IEEEmembership{}and~H\"{u}seyin Arslan,~\IEEEmembership{Fellow,~IEEE}
\thanks{Ahmed T. El-Toukhy is with the Department of Electrical
and Electronics Engineering, Istanbul Medipol University, 34810 Istanbul, Turkey (e-mail: atalaat@st.medipol.edu.tr).}
\thanks{H. Arslan is with the Department of Electrical Engineering, University
of South Florida, Tampa, FL 33620 USA, and also with the Department
of Electrical and Electronics Engineering, Istanbul Medipol University,
34810 Istanbul, Turkey (e-mail: huseyinarslan@medipol.edu.tr).}

}

\maketitle

\begin{abstract}
Cognitive radio networks (CRNs) are considered a promising solution for spectrum resources scarcity and efficient channel utilization. In this letter, multi-dimensional analytical Markov model based on reservation channel access scheme and channel aggregation method is proposed to enhance spectrum utilization, capacity of low priority secondary users (SUs) and  reducing handoff probability of SUs. Moreover, the proposed method improves the performance of high priority SUs by providing the capability to resume the connection after dropping. The numerical results indicate that the modified reservation access model can enhance the performance of SUs compared to the traditional basic random access model.
\end{abstract}


\IEEEpeerreviewmaketitle

\section{Introduction}

\IEEEPARstart{I}{n} the recent years, the fast growth in wireless and communication technologies, combined with the increasing demand for spectral resources, has turned the attention of researchers towards solving spectrum scarcity issue. Cognitive radio (CR) has attained increasing popularity due to its capability of utilizing the idle channels dynamically without affecting the rights of primary user (PU), addressing the spectrum scarcity issue to an extent \cite{akyildiz2006next}, \cite{hu2018full}.

Dynamic channel allocation and spectrum access are identified as a core concepts of CR, where secondary users (SUs) can exploit the idle spectrum of PUs \cite{hu2018full}. In \cite{el2016qos}, a novel channel allocation scheme using Markov model is proposed to boost the performance and quality of service (QoS) for the high priority SU. A dynamic channel aggregation mechanism based on the Markovian prediction of the state of spectrum is presented in \cite{wei2016dynamic} which promises enhanced spectrum efficiency and improved data rate. Continuous time Markov chain (CTMC) models are proposed in \cite{jiao2010analysis} to analyze the performance of the secondary network when channels are opportunistically available for SUs using different channel aggregation or bonding strategies without considering spectral handoff. A dynamic spectrum access and channel reservation Markov model considering the priority of PUs and SUs is proposed to optimize the number of reserved channels  \cite{omer2017utilization}. 

For improving channel utilization, the authors in \cite{piran2016qoe} developed a priority-based spectrum allocation model for SUs based on their quality of experience (QoE) requirements. Also, a management strategy is presented to overcome the interruptions caused by handoff. In \cite{7891952}, a buffering and switching scheme for admission control is proposed to enhance the performance of SUs using CTMC. In \cite{el2017performance}, the authors investigate three dimensional analytical Markov model using random and reservation channel access schemes  considering the priority of SUs to enhance their QoS. In this letter, a novel efficient spectrum resource utilization scheme is proposed using analytical five dimensional Markov model based on channel aggregation and dynamic reservation channel access assignment to enhance the performance of SUs. Furthermore, the complexity analysis of the proposed model is provided.

Thus, comparing with the work in \cite{el2017performance}, our contribution can be summarized as follows:

\begin{itemize}
\item Minimizing starvation of low priority SUs.
\item Improving spectrum utilization and capacity of low priority SUs. 
\item Decreasing the handoff probability of SUs.
\item Improving the performance of high priority SUs by providing the capability to resume the connection after dropping.
\end{itemize}
Increasing number of channels in this work enabled us to implement the aggregation and reservation channel model.

This letter is organized as follows: Section II describes the analytical models; Section III presents the performance metrics and complexity analysis; Section IV provides numerical results and analysis;  Section V concludes the letter.

\begin{table*}[thb]

\centering
\caption {State transitions of basic random access model at state $S(i,j_1,j_2).$}
\resizebox{1\textwidth}{!}{
\begin{tabular}{l l l c}
\hline
\textbf {Event / Action} & \textbf {Conditions} & \textbf {New state} & \textbf {Trans rate} \\ 
\hline

AR. PU / PU will utilize idle channel. & \ $N^{idle} > 0$  &\ $S' (i+1,j_1,j_2)$ &\ $a^*$\\
\hline
\multirow{2}{*}{AR. PU / SU will handoff to idle channel.} & \ $N^{idle} > 0$, PU accesses&\ \multirow{2}{*}{$S' (i+1,j_1,j_2)$} &\ \multirow{2}{*}{ $b^*$}\\
& channel utilized by SU.  & & \\
\hline
AR. PU / One of low priority SU-2 will be dropped. & \ $N^{idle} = 0$, $j_2 > 0$  &\ $S' (i+1,j_1,j_2-1)$ &\ $c^*$\\
\hline

AR. PU / One of high priority SU-1 will be dropped. & \ $N^{idle} = 0$, $j_1 > 0$, $j_2 = 0 $  &\ $S' (i+1,j_1-1,j_2)$ &\ $c^*$\\
\hline

Dep. PU / Arrival PU/SU can use this idle channel. & \ $----$  &\ $S' (i-1,j_1,j_2)$ &\ $i\mu_p$\\
\hline

AR. SU-1 / SU-1 will occupy any idle channel. & \ $N^{idle} > 0$&\   $S' (i,j_1+1,j_2)$ &\ $\lambda_s$\\

\hline
AR. SU-1 / One of the low priority SUs-2 will be dropped. & \ $N^{idle} = 0$, $j_2 > 0$  &\ $S' (i,j_1+1,j_2-1)$ & \ $\lambda_s$\\
\hline
AR. SU-1 / SU-1 will be blocked. & \ $N^{idle} = 0$, $j_2 = 0$ &\ $----$ &\ $----$ \\
\hline

Dep. SU-1 / Arrival PU/SU can use this idle channel. & \ $----$   &\ $S' (i,j_1-1,j_2)$ &\ $j_1\mu_s$ \\
\hline

AR. SU-2 / SU-2 will occupy any idle channel. & \ $N^{idle} > 0$  &\ $S' (i,j_1,j_2+1)$ &\ $\lambda_s$ \\

\hline
AR. SU-2 / SU-2 will be blocked. & \ $N^{idle} = 0$  &\ $----$ &\ $----$ \\
\hline

Dep. SU-2 / Arrival PU/SU can use this idle channel. & \ $----$  &\ $S' (i,j_1,j_2-1)$ &\ $j_2 \mu_s$ \\
\hline
\
*$a=\frac{(M-C_x)}{(M-i)}(k-i)\lambda_p$
& *$b=\frac{(j_1+j_2)}{(M-i)}(k-i)\lambda_p$
& *$c=(k-i)\lambda_p$

\end{tabular}
}

\end{table*}

\section{Analytical Models}

Cognitive radio model consists of a central base station (CBS), finite PUs and infinite SUs, where PUs and SUs are operating on the same licensed spectrum
band which is divided into $M$ homogeneous channels. Furthermore, $k$ represents the total number of PUs and equal bandwidth channels are assumed for each SU and PU. PUs have the right to use and reclaim channels at any time, but each PU can only occupy one channel in its service. The arrival rates of SUs and PUs follow the Poisson process with rates of $\lambda_s\ $and $\lambda_p\ $, respectively. The service times of SUs and PUs follow negative exponential distribution with service rates $\mu_s\ $and $\mu_p\ $, respectively. CBS has historical database of QoS which is collected by perfect spectrum sensing of SUs. Therefore, CBS is responsible to allocate suitable idle channels to SUs and classify SUs and channels based on channel utilization.

\subsection{Basic Random Access Model}
In this model, it is assumed that SUs are classified to $l$ different classes of priorities where the element $j_l$ represents the number of SUs of class $l$ $(l=1,2)$. Random access of channel due to calls follows CTMC model. The states of Markov model are represented as $S(i,j_1,j_2)$, where $i,j_1$ and $j_2$ represents the number of primary channels occupied by PUs, the number of high priority class-1 SUs (SUs-1) and the number of low priority class-2 SUs (SUs-2), respectively. $M$ is the total number of available channels. Furthermore, $C_x=(i+j_1+j_2)$ and $N^{idle} = M-C_x$ are the total number of occupied and idle channels at any state, respectively. The state of Markovian model $S(i,j_1,j_2)$ will be changed to another state $S'(i,j_1,j_2)$ based on arrival/departure of PUs and SUs as summarized in TABLE I. The steady state probability vector $\pi$ can be calculated by solving linear equation $\pi \textbf{Q}=0$ under the constraint $\pi \textbf{e}=1$ using numerical method presented in \cite{cooper1981introduction}, where $\textbf{Q}$ is the transition rate matrix for this basic Markovian model and $\textbf{e}$ is a column vector with all ones.

\subsection{Proposed Modified Reservation Access Model}
The proposed method improves the basic model by supporting the low priority SUs-2 to occupy more channels, minimize their starvation of resources and enhance the performance by using reservation channels access and channels aggregation method. This strategy allows the channels to be grouped and specifically used for SUs-2. Moreover, the proposed method also enhances the performance of SUs-1, especially urgent data users by allowing them to resume the connection after it is dropped. In order to achieve this, SUs are divided into three different classes based on their priorities. 

CTMC model is proposed to describe the reservation channel access strategy. The state of modified Markov model is represented as $Z(i,j'_1,j_1,j_m,j_n)$, where $i$ represents the number of primary channels occupied by PUs, $j'_1, j_1, j_m$ and $j_n$ represent the number of returned-class-1 (SUs-R1),i.e, urgent data SUs who resume the connection after dropping, the number of SUs-1, i.e, real time data SUs, the number of SUs-2, i.e, non-real time data SUs who aggregate $m$ and $n$ channels, respectively. $m$ and $n$ represent the maximum and minimum number of aggregation channels, respectively. Consequently, the number of SUs-2 is $j_2=( j_m + j_n)$. Any incoming PU should first occupy $M_{rp}$ which are reserved channels for PUs. Afterwards, PUs will start to occupy unreserved channels randomly.

Assume that SUs-R1 have the ability to utilize $M'_1$ channels only. However, if SUs-R1 do not exist, SUs-1 can utilize these channels. $M_{r2}$ is the number of reserved channels for SUs-2 only. Thus, the total available number of channels for SUs-1 and SUs-2 is $M_1=(M-M_{rp}-M_{r2})$ and $M_2=(M-M_{rp}-M'_1)$, respectively. While, $M_x=[i+j'_1+j_1+mj_m+nj_n]$ and $M^{idle}=(M-M_x)$ are the total number of occupied and idle channels at any state, respectively. According to arrival/departure of PUs and SUs, the state transition of Markov model $Z(i,j'_1,j_1,j_m,j_n)$ will be changed to another state $Z'(i,j'_1,j_1,j_m,j_n)$ as summarized in TABLE II. The steady state probability vector $\pi$ can be calculated by solving linear equation $\pi \textbf{P}=0$ under the constraint $\pi \textbf{e}=1$ using numerical method presented in \cite{cooper1981introduction}, where $\textbf{P}$ denotes the transition rate matrix for Markov model and $\textbf{e}$ is a column vector with all ones.

\section{Performance Metrics and Complexity Analysis}

In this section, the evaluation metrics equations will be derived. The performance of the basic and proposed modified models will be evaluated in terms of spectrum utilization, capacity, blocking probability and hand off probability of SUs.

\begin{table*}[htbp]
\centering
\caption {State transitions of modified reservation access model at state $Z(i,j'_1,j_1,j_m,j_n).$}
\resizebox{1\textwidth}{!}{  
\begin{tabular}{l l l c}
\hline
\textbf {Event / Action} &  \textbf {Conditions} & 
\textbf {New state} &  \textbf {Trans rate}  \\ 
\hline
AR. PU / PU will utilize reserved idle channel.& \ $M^{idle} > 0$, $i< M_{rp} $  &\ $Z' (i+1,j'_1,j_1,j_m,j_n)$ &\ $c'^*$\\
\hline
AR. PU / PU will utilize unreserved idle channel. & \ $M^{idle} > 0$, $M_{rp}\leq i< M $   &\ $Z' (i+1,j'_1,j_1,j_m,j_n)$ &\  $a'^*$\\
\hline
\multirow{2}{*}{AR. PU / SU will handoff to another idle channel.} & \ $M^{idle} > 0$, $M_{rp}\leq i< M $, PU&\ \multirow{2}{*}{$Z' (i+1,j'_1,j_1,j_m,j_n)$} &\ \multirow{2}{*}{$b'^*$}\\
& accesses channel utilized by SU.  &  & \\
\hline
AR. PU / One of  aggregated SUs-2 will lose & \ \multirow{2}{*}{$M^{idle} = 0$, $j_m>0 $}  &\ \multirow{2}{*}{$Z' (i+1,j'_1,j_1,j_m-1,j_n+1)$} &\ \multirow{2}{*}{$c'^*$}\\
one of his aggregated channels.& & \\
\hline
\multirow{2}{*}{AR. PU / One of SUs-R1 will be dropped.} & \ $M^{idle} = 0$, $j_m=0 $, PU accesses&\ \multirow{2}{*}{$Z' (i+1,j'_1-1,j_1,j_m,j_n)$} &\ \multirow{2}{*}{$c'^*$}\\
& channel utilized by SU-R1. & & \\
\hline
\multirow{2}{*}{AR. PU / One of SUs-1 will be dropped.} & \ $M^{idle} = 0$, $j_m=0 $, PU accesses&\ \multirow{2}{*}{$Z' (i+1,j'_1,j_1-1,j_m,j_n)$} &\ \multirow{2}{*}{$c'^*$}\\
& channel utilized by SU-1. & & \\
\hline
\multirow{2}{*}{AR. PU / One of SUs-2 will be dropped.} & \ $M^{idle} = 0$, $j_m=0 $, PU accesses&\ \multirow{2}{*}{$Z' (i+1,j'_1,j_1,j_m,j_n-1)$} &\ \multirow{2}{*}{$c'^*$}\\
& channel utilized by SU-2. & & \\
\hline
Dep. PU / Arrival PU/SU can use this idle channel. & \ $----$  &\ $Z' (i-1,j'_1,j_1,j_m,j_n)$ &\ $i\mu_p$\\
\hline

AR. SU-R1 / SU-R1 will occupy any idle channel. & \ $M^{idle} > 0$, $j'_1< M'_1 $ &\ $Z' (i,j'_1+1,j_1,j_m,j_n)$ &\ $\lambda_s$\\
\hline
AR. SU-R1 / One of aggregated SUs-2 will lose & \ \multirow{2}{*}{$M^{idle} = 0$, $j'_1< M'_1$, $j_m>0$} &\ \multirow{2}{*}{$Z' (i,j'_1+1,j_1,j_m-1,j_n+1)$} &\ \multirow{2}{*}{$\lambda_s$}\\
one of his aggregated channels.& & \\
\hline
\multirow{2}{*}{AR. SU-R1 / One of SUs-2 will be dropped.} & \ $M^{idle} = 0$, $j'_1< M'_1 $, $j_m=0$,&\ 
\multirow{2}{*}{$Z' (i,j'_1+1,j_1,j_m,j_n-1)$} &\ \multirow{2}{*}{$\lambda_s$}\\
& $j_n>M_{r2}$ & & \\
\hline
\multirow{2}{*}{AR. SU-R1 / One of SUs-1 will be dropped.} & \ $M^{idle} = 0$, $j'_1< M'_1 $, $j_1>0$,  &\ \multirow{2}{*}{$Z' (i,j'_1+1,j_1-1,j_m,j_n)$} &\ \multirow{2}{*}{$\lambda_s$}\\
& $j_m=0$, $j_n\leq M_{r2}$ & & \\
\hline
\multirow{2}{*}{AR. SU-R1 / SU-R1 will be blocked.} & \ $M^{idle} = 0$, $j'_1 \leq M'_1 $, $j_1=0$, &\ \multirow{2}{*}{$----$} &\ \multirow{2}{*}{$----$}\\
& $j_m=0$, $j_n\leq M_{r2}$ & & \\
\hline
Dep. SU-R1 / Arrival PU/SU can use this idle channel. & \ $----$ &\ $Z' (i,j'_1-1,j_1,j_m,j_n)$ &\ $j'_1\mu_s$\\
\hline

AR. SU-1 / SU-1 will occupy any idle channel. & \ $M^{idle} > 0$, $j_1< M_1 $ &\ $Z' (i,j'_1,j_1+1,j_m,j_n)$ &\ $\lambda_s$\\
\hline
AR. SU-1 / One of aggregated SUs-2 will lose & \
\multirow{2}{*}{$M^{idle} = 0$, $j_1< M_1 $, $j_m>0$} &\ \multirow{2}{*}{$Z' (i,j'_1,j_1+1,j_m-1,j_n+1)$}& \ \multirow{2}{*}{$\lambda_s$}\\
one of his aggregated channels.& & \\
\hline
\multirow{2}{*}{AR. SU-1 / One of SUs-2 will be dropped.} & \ $M^{idle} = 0$, $j_1< M_1 $, $j_m=0$, &\ \multirow{2}{*}{$Z' (i,j'_1,j_1+1,j_m,j_n-1)$} &\ \multirow{2}{*}{$\lambda_s$}\\ & $j_n>M_{r2}$& & \\
\hline
AR. SU-1 / SU-1 will be blocked. & \ $M^{idle} = 0$, $j_m=0$, $j_n\leq M_{r2}$  &\ $----$ &\ $----$\\
\hline
Dep. SU-1 / Arrival PU/SU can use this idle channel & \ $----$ &\ $Z' (i,j'_1,j_1-1,j_m,j_n)$ &\ $j_1\mu_s$\\
\hline

AR. SU-2 / SU-2 will aggregate $m$ idle channels. & \ $M^{idle} > 0$, $(mj_m+nj_n)<M_2$ &\ 
$Z' (i,j'_1,j_1,j_m+1,j_n)$ &\ 
$\lambda_s$\\

\hline
AR. SU-2 / One of aggregated SUs-2 will lose & \ $M^{idle} \geq 0$, $j_m>0 $, $j_n \geq 0$,  &\ \multirow{2}{*}{$Z' (i,j'_1,j_1,j_m-1,j_n+2)$} &\ \multirow{2}{*}{$\lambda_s$}\\
one of his aggregated channels.& 
$(mj_m+nj_n)<M_2$ & & \\
\hline
AR. SU-2 / SU-2 will be blocked. & \ $M^{idle} = 0$, $j_m=0$, $j_n=M_2$  &\ $----$ &\ $----$\\
\hline
Dep. SU-2 / Arrival PU/SU can use this idle channel. & \ $j_m>0$, $j_n=0$  &\ $Z' (i,j'_1,j_1,j_m-1,j_n)$ &\ $j_2\mu_s$\\
\hline
Dep. SU-2 / Arrival PU/SU can use this idle channel. & \ $j_m \geq 0$, $j_n>0$ &\ $Z' (i,j'_1,j_1,j_m+1,j_n-2)$ &\ $j_2\mu_s$\\
\hline

*$a'=\frac{M-(i+j'_1+j_1+j_2)}{(M-i)} (k-i)\lambda_p$
& *$b'=\frac{(j'_1+j_1+j_2)}{(M-i)} (k-i)\lambda_p$
& *$c'=(k-i)\lambda_p$

\end{tabular}
}
\end{table*}

\subsection{Capacity}

The capacity of SUs is denoted as the average number of completed service requests per unit time. In other word, the capacity is defined as the multiplication of the total number of SU, service rate of SU and steady state probability of SU at the target states \cite{ngatched2013analysis}.

\subsubsection{Basic random channel access model} The capacity of SUs can be expressed as $\rho_1$ and $\rho_2$ for SUs-1 and SUs-2, respectively.
\begin{equation}
\rho_1=\sum_{s\in S}^{} j_1\mu_s\pi_s.
\end{equation}

\vspace{-1mm} 

\begin{equation}
\rho_2=\sum_{s\in S}^{} j_2\mu_s\pi_s.
\end{equation}

\subsubsection{Modified reservation channel access model} The capacity of SUs can be expressed as $\rho'_{R1}$, $\rho'_1$ and $\rho'_2$ for SU-R1, SUs-1 and SUs-2, respectively.

\vspace{-3mm}

\begin{equation}
\rho'_{R1}=\sum_{z\in Z}^{} j'_1\mu_s\pi_z.
\end{equation}

\vspace{-2mm}

\begin{equation}
\rho'_1=\sum_{z\in Z}^{} j_1\mu_s\pi_z.
\end{equation}

\vspace{-2mm}

\begin{equation}
\rho'_2=\sum_{\substack{k=n,z\in Z}}^{m} kj_k\mu_s\pi_z.
\end{equation}

\subsection{Spectrum Utilization}

The spectrum utilization of the network can be  defined as the ratio between the average number of occupied channels and total number of channels at the target states \cite{zukerman2013introduction}.

\subsubsection{Basic random channel access model} The spectrum utilization, $U$, is given by

\vspace{-2.5mm}

\begin{equation}
U=\sum_{s\in S}^{} \frac{C_x}{M}\pi_s.
\end{equation}

\subsubsection{Modified reservation channel access model} The spectrum utilization, $U'$, is given by

\vspace{+.3mm}

\begin{equation}
U'=\sum_{z\in Z}^{} \frac{M_x}{M}\pi_z.
\end{equation}

\subsection{Blocking Probability}

If a system is totally occupied, the arriving SU will be blocked from getting any resources. The  probability of this event is presented in \cite{zukerman2013introduction,5285171,5633740} and evaluated by

\begin{equation}
P_b = \frac{\text{\ Total SU blocking rate}}{\text{\ Total user arrival rate}}=\frac{\lambda_s\pi}{(k-i)\lambda_p+\lambda_s}.
\end{equation}

\subsubsection{Basic random channel access model}Let $P_{b1}$ and $P_{b2}$ be the blocking probability of SU-1 and SU-2, respectively, \cite{el2017performance}.

\vspace{-5mm}

\begin{equation}
P_{b1}=\sum_{\substack{{i=0}, s\in S}}^M\sum_{\substack{j_1=0,j_2=0;N^{idle}=0}}^{M} \frac{\lambda_s\pi_s}{(k-i)\lambda_p+\lambda_s}.
\end{equation}

\vspace{+1mm}

\begin{equation}
P_{b2}=\sum_{\substack{{i=0},s\in S}}^M\sum_{\substack{j_1=0,N^{idle}=0}}^{M-(i+j_2)} \frac{\lambda_s\pi_s}{(k-i)\lambda_p+\lambda_s}.
\end{equation}

\subsubsection{Modified reservation channel access model}Let $P'_{bR1}$, $P'_{b1}$ and $P'_{b2}$ be the blocking probability of SU-R1, SU-1 and SU-2, respectively.

\vspace{-3mm}

\begin{equation}
P'_{bR1}=\sum_{\substack{i=0,z\in Z;j'_1=M'_1}}^{M} \frac{\lambda_s\pi_z}{(k-i)\lambda_p+\lambda_s}.
\end{equation}

\vspace{-3mm}

\begin{equation}
P'_{b1}=\sum_{\substack{i=0,z\in Z}}^{M}\sum_{\substack{j_1=0,j'_1+j_1=M_1;\\i+j'_1+j_1\geq M-M_{r2}}}^{M_1} \frac{\lambda_s\pi_z}{(k-i)\lambda_p+\lambda_s}.
\end{equation}

\vspace{-3mm}

\begin{equation}
P'_{b2}=\sum_{\substack{i=0,z\in Z}}^{M}\sum_{\substack{j_1=0,j_n=M_2,j_m=0;\\j'_1+j_1+j_n=M_2,j_m=0}}^{M_1} \frac{\lambda_s\pi_z}{(k-i)\lambda_p+\lambda_s}.
\end{equation}

\subsection{Hand off Probability}
If PUs arrive to a certain channel which is occupied by SU while idle channels are available, SU will be handed off to the idle channel to resume the transmission. The handoff probability is presented in \cite{5633740} and evaluated by

\vspace{-4mm}

\begin{equation}
P_h=\frac{\text{\ Total SU transition rate}}{\text{\ Total user connection rate}}=\frac{\frac{f(j)}{M-i}(k-i)\lambda_p\pi}{f(P_b) ((k-i)\lambda_p+\lambda_s)},
\end{equation}

\noindent where $f(j)$ and $f(P_b)$ are two factors that depend on the number of the existing SUs at the state and blocking probability, respectively, and vary based on the class of SU.\\

\subsubsection{Basic random channel access model}Let $P_{h1}$ and $P_{h2}$ be the handoff probability of SU-1 and SU-2, respectively \cite{el2017performance}.
 
\begin{equation}
P_{h1}=\hspace{-2mm}\sum_{\substack{{i=0},s\in S}}^M\sum_{\substack{j_1=1,N^{idle}>0}}^{M-(i+j_2+1)} \frac{\frac{j_1}{M-i}(k-i)\lambda_p\pi_s}{(1-P_{b1})((k-i)\lambda_p+\lambda_s)}.
\end{equation}

\begin{equation}
P_{h2}=\hspace{-2mm}\sum_{\substack{{i=0},s\in S}}^M\sum_{\substack{j_2=1,N^{idle}>0}}^{M-(i+j_1+1)} \frac{\frac{j_1+j_2}{M-i}(k-i)\lambda_p\pi_s}{(1-P_{b2})((k-i)\lambda_p+\lambda_s)}.
\end{equation}

\subsubsection{Modified reservation channel access model}
Let $P'_{hR1}$, $P'_{h1}$ and $P'_{h2}$ be the handoff probability of SU-R1, SU-1 and SU-2, respectively.

\vspace{-4mm}

\begin{equation}
P'_{hR1}=\hspace{-2mm}\sum_{\substack{i=M_{rp},z\in Z;j'_1=M'_1\\M_x=M,j_m>0;\\M_x<M,j_m=j_n=0;\\M_x<M,j_n=0}}^{M}\frac{\frac{j'_1}{(M-i)}(k-i)\lambda_p \pi_z}{(1-P'_{bR1})((k-i)\lambda_p+\lambda_s)}.
\end{equation}
 
\vspace{-3mm} 
 
\begin{equation}
P'_{h1}=\hspace{-3mm}\sum_{\substack{i=M_{rp}\\z\in Z}}^{M}\sum_{\substack{j_1=1\\M_x<M,j_n=0\\M_x=M,j_m>0\\M_x<M,j_m=j_n=0}}^{M-M'_1}\frac{\frac{j'_1+j_1}{(M-i)}(k-i)\lambda_p \pi_z}{(1-P'_{b1})((k-i)\lambda_p+\lambda_s)}.
\end{equation}

\vspace{-3mm}

\begin{equation}
P'_{h2}=\hspace{-2mm}\sum_{\substack{i=M_{rp},z\in Z\\M_x=M,j_2\neq0,j_m>0\\M_x<M,j_n=0,j_m\neq0;\\M_x<M,j_m=0,j_n\neq0;\\M_x<M,j_m\neq0,j_n\neq0}}^{M}\frac{\frac{j'_1+j_1+j_2}{(M-i)}(k-i)\lambda_p \pi_z}{(1-P'_{b2})((k-i)\lambda_p+\lambda_s)}.
\end{equation}

\subsection{Complexity Analysis}

In this subsection, the complexity of the basic and proposed modified models is measured by the number of total states that represents both models tacking into account the state dimension size. For the basic model, the states are represented by a 3-D Markov model. Thus, the complexity of basic model, $\psi_{basic}$, is given by

\vspace{-5.5mm}

\begin{equation}
\psi_{basic}=\sum_{\substack{v=1}}^{M+1}\sum_{\substack{w=1}}^{v} w=\frac{M^{3}}{6}+M^{2}+\frac{11 M^{}}{6}+1.
\end{equation}

\noindent For the proposed model, the states are represented by a 5-D Markov model. Therefore, the complexity of proposed model, $\psi_{prop}$, is given by

\vspace{-7.5mm}

\begin{equation} \label{pdfOFDMHNIM}
\begin{split}
\hspace{-.5mm} \psi_{prop}=(M_{rp}+1)\left(\sum_{\substack{v=M_{rp}}}^{M-M_{rp}}2v+(M-M_{rp})\right)+\hspace{-2mm}\sum_{\substack{w=1}}^{M-M_{rp}}w^{2} \\ &  \hspace{-45mm} =\frac{M^{3}}{3}+\frac{3 M^{2}}{2}-\frac{23 M^{}}{6}-7.
\end{split}
\end{equation}


\vspace{-1mm}

\noindent It should be noted that the last line in (21) is calculated at $M_{rp}=2$ as used in the proposed model. From (20) and (21), the complexity order of the basic and proposed models are $\mathcal{O}(\frac{M^{3}}{6})$ and $\mathcal{O}(\frac{M^{3}}{3})$, respectively.\\


\section{Numerical Results and Analysis}

In this section, the performance of the basic and modified models will be analyzed with different arrival and service rates of SUs in terms of spectrum utilization, capacity, blocking probability and handoff probability using the derived analytical equations. The parametric values of analytical models are set as $M$=7, $M_{rp}$=2, $M'_1$=1, $M_{r2}$=1, $m$=2, $n$=1, $k$=10, $\lambda_p$=0.05 call/sec and $\mu_p$=0.4 call/sec. In case of increasing SUs' arrival rate values used are $\lambda_s$=0.1 to 0.5 call/sec and $\mu_s$=0.5 call/sec. Similarly, for increasing SUs' service rate values are set as $\lambda_s$=0.25 call/sec and $\mu_s$=0.25 to 0.5 call/sec.

As shown in Fig. 1(a), the capacity of SUs increases with SUs' arrival rates because of increasing channel access requests. Furthermore, the modified model increases the capacity of SUs-2 as compared to the basic model due to channel reservation strategy. In contrast, the capacity of SUs-R1 is small because of their ability to access only one channel to continue the connection. Fig. 1(b) illustrates that, spectrum utilization is improved for the modified model due to the applied aggregation method of channels which enables efficient utilization. In addition, this figure indicates that increasing SUs' arrival rate leads to reduced idle time, improving spectrum utilization. 

\begin{figure}[htbp]\centering
\centerline{\includegraphics[width=1.09\columnwidth]{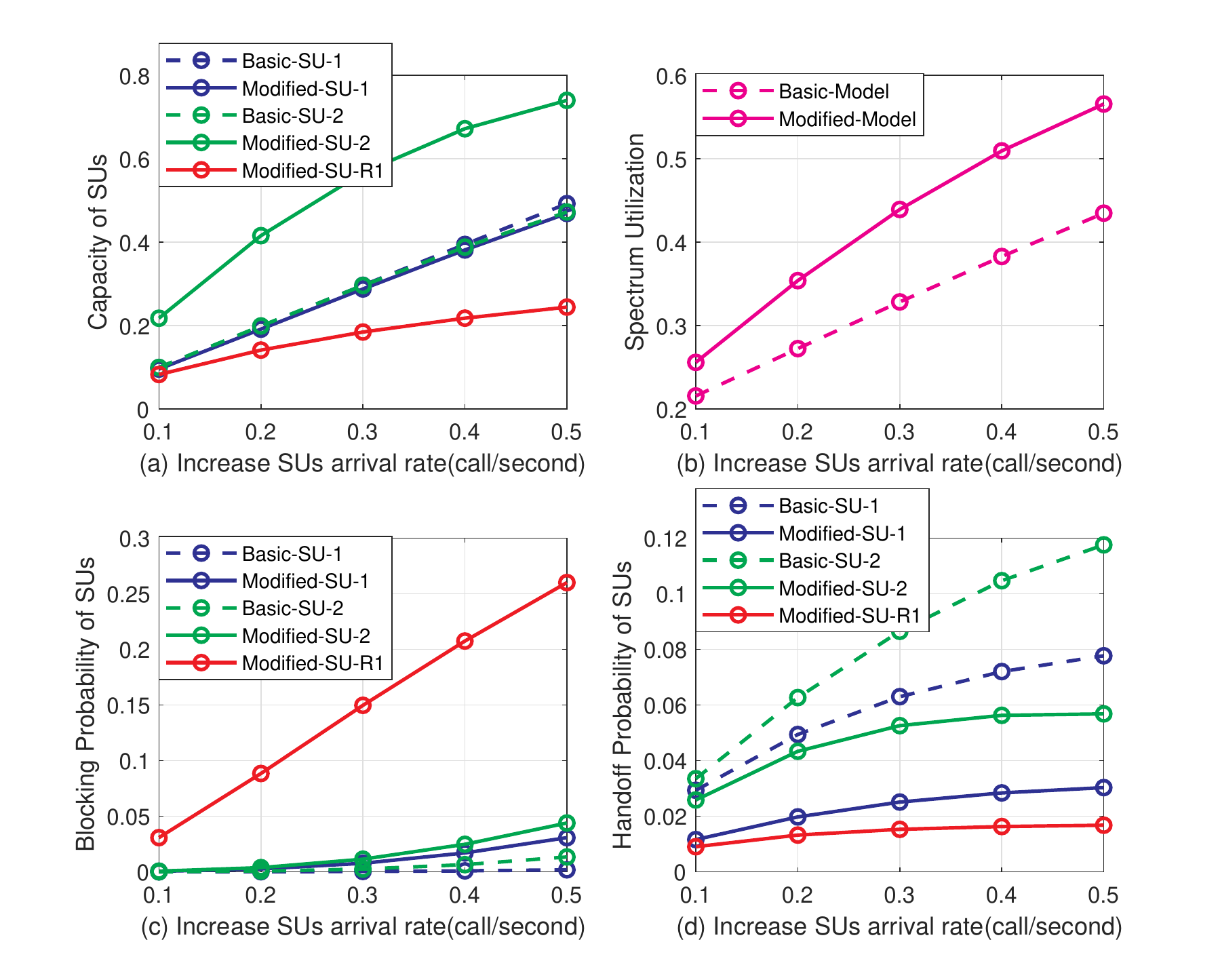}}
\caption{Performance metrics of SUs for SUs' arrival rates.}
\label{fig}
\end{figure}

\begin{figure}[htbp]\centering
\centerline{\includegraphics[width=1.09\columnwidth]{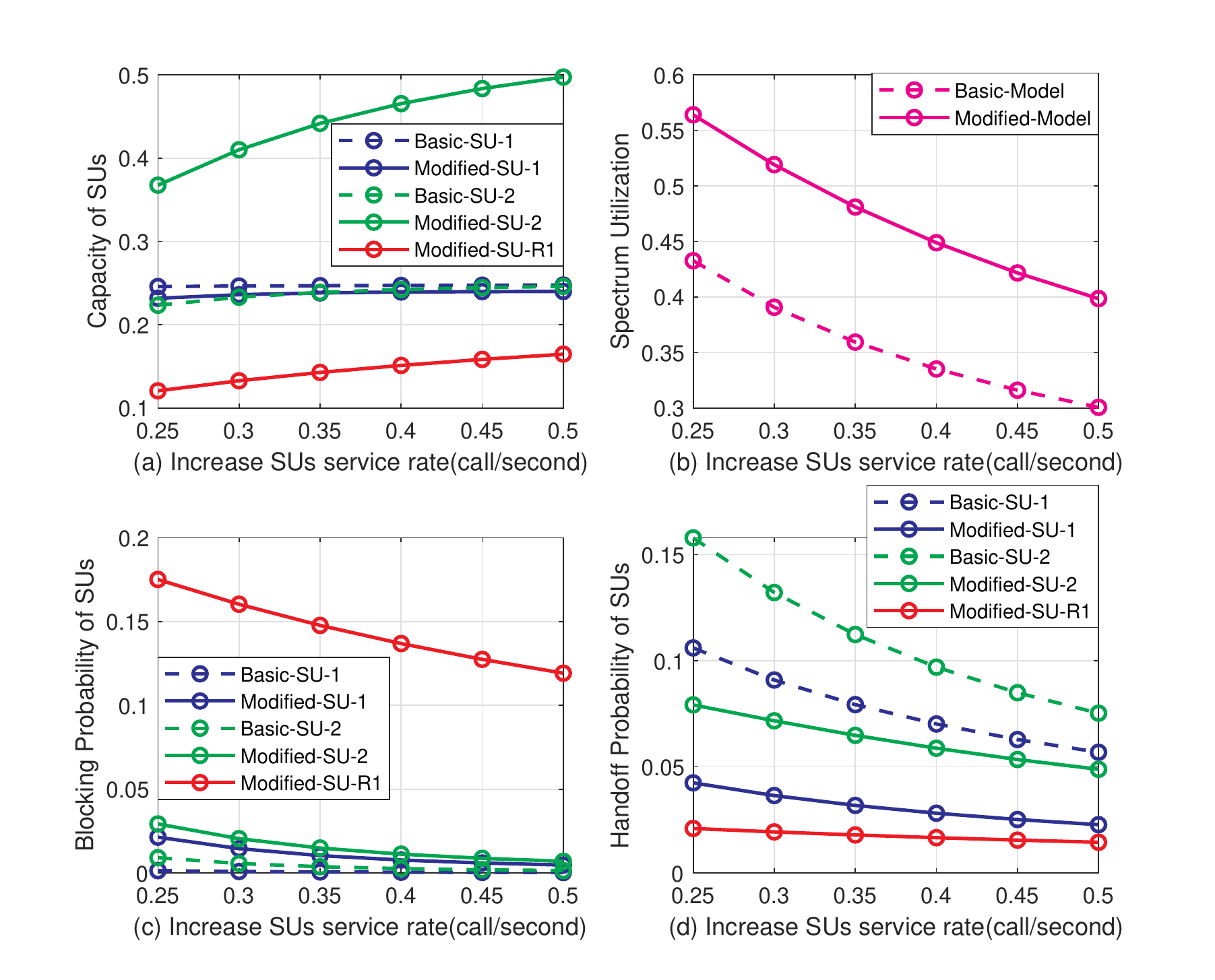}}
\caption{Performance metrics of SUs for SUs' service rates.}
\label{fig}
\end{figure}

Fig. 1(c) illustrates that the blocking probability of SUs increases with increasing SU arrival rate because of the increased channel access requests. Moreover, SUs-R1 have the higher blocking probability because of the limited number of channels which are available for them. Fig. 1(d) shows reduction in handoff probability for the modified model comparing to the basic model due to reserved channels for PUs and SUs-2. Accordingly, the probability that PUs access certain channels occupied by SUs is reduced. In addition, this figure shows that increasing SUs' arrival rate will increase handoff probability because of higher channel access request.

In Fig. 2(a), idle time will increase with the increase in service rate of SUs, i.e., more channels will be available for the different classes of SUs. Consequently, the capacity of SUs is increased. Also, it is noticeable that low priority SUs-2 in the modified model have the highest capacity compared to the basic model due to reserved channels while SUs-R1 have the lowest capacity because of the limitation of their available channels. Fig. 2(b) shows that the modified model enhances the spectrum utilization compared to the basic model because of using channel aggregation ensuring more efficient utilization of channels. The decreasing trend of the same with increase in service rate is due to the fact that with this faster service, channels will be idle for longer periods of time. 
Due to the same reason, blocking probability will be reduced, as shown in Fig. 2(c). Following the same reasoning, there would be lesser number of handoffs required, as illustrated in Fig. 2(d). Note that handoff probability of the modified model is less than that of the basic model for different classes of SUs as a direct result of using reserved channels.

\section{Conclusion}

In this letter, a novel efficient Markov model based spectrum utilization scheme is proposed which enables dynamic reserved channel access and channel aggregation. The numerical results of the modified model show significant improvement in the performance of cognitive radio system compared to the basic model by minimizing  starvation of low priority SUs, improving spectrum utilization and capacity of low  priority  SUs, decreasing the handoff probability of SUs and enabling high priority SUs to resume the connection after dropping at the cost of increased blocking probability.

\section*{Acknowledgment}

The authors would like to thank Muhammad Sohaib J. Solaija for his fruitful comments regarding this manuscript.

\ifCLASSOPTIONcaptionsoff
  \newpage
\fi

\bibliographystyle{IEEEtran} 

\end{document}